\ifx \doiurl    \undefined \def \doiurl#1{\href{http://dx.doi.org/#1}{\textsf{DOI}}}\fi
\ifx \adsurl    \undefined \def \adsurl#1{\href{http://adsabs.harvard.edu/abs/#1}{\textsf{ADS}}}\fi

\documentclass[namedreferences]{solarphysics}
\usepackage[]{spr-sola-addons} 
\usepackage[english]{babel}
\usepackage{graphicx}
\usepackage{natbib}
\usepackage{float}
\usepackage{savesym}
\usepackage{bm}
\usepackage{color}          
\usepackage{breakurl}       
\usepackage{txfonts}
\usepackage{wasysym}
          
\newcommand {\rsun}{R$_{\odot}$}

\newcommand{\fe}{Fe~{\sc{xiv}}}
\newcommand{\dg}{$^{\circ}$}
\newcommand{\arcsec}{"}

\newcommand{\aap}{    {\it Astron. Astrophys.}}

\newcommand{\apj}{    {\it Astrophys. J.}}
\newcommand{\apjl}{   {\it Astrophys. J. Lett.}}

\newcommand{\grl}{    {\it Geophys. Res. Lett.}}

\newcommand{\jastp}{  {\it J. Atmos. Solar-Terr. Phys.}} 
\newcommand{\jgr}{    {\it J. Geophys. Res.}}
\newcommand{\mnras}{  {\it Mon. Not. Roy. Astron. Soc.}}
\newcommand{\nat}{    {\it Nature}}

\newcommand{\pre}{    {\it Phys. Rev. E}}
\newcommand{\solphys}{{\it Solar Phys.}}
 
\newcommand{\ssr}{    {\it Space Sci. Rev.}} 
\chardef\us=`\_

\begin{document}

\begin{article}
\begin{opening}

\title{Spatio--Temporal Evolution and North--South Asymmetry of Quasi-Biennial Oscillations in the Coronal \fe\ Emission}

\author[addressref=aff1,corref,email={salvatore.mancuso@inaf.it}]{\inits{S.}\fnm{S.}~\lnm{Mancuso}}
\author[addressref=aff2,]{\inits{T.S.}\fnm{T. S.}~\lnm{Lee}}
\author[addressref={aff1,aff3}]{\inits{C.}\fnm{C.}~\lnm{Taricco}}
\author[addressref={aff1,aff3}]{\inits{S.}\fnm{S.}~\lnm{Rubinetti}}
                   
\address[id=aff1]{Istituto Nazionale di Astrofisica, Osservatorio Astrofisico di Torino, Strada Osservatorio 20, 10025 Pino Torinese, Italy}
\address[id=aff2]{School of Science and Technology, Georgia Gwinnett College, Lawrenceville, GA, USA}
\address[id=aff3]{Dipartimento di Fisica, Universit\`a degli Studi di Torino, Via P. Giuria 1, 10125 Torino, Italy}

\runningtitle{Spatio--Temporal Evolution and North--South Asymmetry of QBOs in Coronal \fe\ Emission}
\runningauthor{S. Mancuso {\it et al.}}

\begin{abstract}
In this work, we apply multichannel singular spectrum analysis (MSSA), a data-adaptive, multivariate, non-parametric technique that simultaneously exploits the spatial and temporal correlations of the input data to extract common modes of variability, to investigate the intermediate quasi-periodicities of the \fe\ green coronal emission line at 530.3 nm for the period between 1944 and 2008. 
Our analysis reveals several significant mid-term periodicities in a range from about one to four years that are consistent with the so-called quasi-biennial oscillations (QBOs), which have been detected by several authors using different data sets and analysis methods.
These QBOs display amplitudes varying significantly with time and latitude over the six solar cycles (18 to 23) covered by this study. 
A clear North--South asymmetry is detected both in their intensity and period distribution, with a net predominance of spectral power in the active-region belt of the northern hemisphere. On the other hand, while the QBOs with periods $\gtrsim 1.7$ years are particularly intense around the polar regions and therefore related to the global magnetic field, the ones with shorter periods are mainly generated at mid-latitudes, in correspondence with the emergence of active regions. Our findings indicate that the North--South asymmetry manifested in the uneven latitudinal distribution of QBOs is a fundamental, albeit puzzling, characteristic of solar activity. 
\end{abstract}
\end{opening}

\section{Introduction}

Solar activity fluctuates with time, exhibiting a wide range of periodicities from scales of a few minutes up to thousands of years as evinced from direct measurements and from proxies based on cosmogenic isotopes.
Except for the nearly periodic 11-year activity cycle, due to the emergence of sunspots in a butterfly-shaped pattern, and the 27-day synodic period, due to solar rotation, the question of the origin and evolution of all other periodicities is still under debate.
Among the latter, variations at timescales between about one and four years, the so-called quasi-biennial oscillations (QBOs), and variations at timescales of several months, the so-called Rieger-type periodicities (Rieger {\it et al.}, 1984), are gaining increasing attention.
These oscillatory modes appear ubiquitous in observations pertaining to the Sun and have been detected in activity proxies that are sensitive to the solar interior, the solar atmosphere, the corona, and even the interplanetary medium (see Bazilevskaya {\it et al.}, 2014 for a recent review). 

The 11-year sunspot cycle is attributed to the large-scale solar dynamo mechanism operating in the solar interior, but the physical reason for the occurrence of these shorter periodicities, also referred to as intermediate- or mid-term quasi-periodicities, is not completely clear. 
In this respect, a number of candidate mechanisms have been proposed in the literature.
One plausible explanation for their origin involves the possible presence of two different dynamo processes acting in the deep and near-surface layers of the convective zone that are responsible, respectively, for the sunspot cycle and the shorter variations (see, {\it e.g.}, Benevolenskaya, 1998; Fletcher {\it et al.}, 2010; Obridko and Badalyan, 2014; Beaudoin {\it et al.}, 2016).
Indeed, periodic variations of 1.3 years in the differential rotation of the deep interior have been discovered by helioseismology (Howe {\it et al.}, 2000) suggesting that the QBOs might be sub-surface in origin.
Further support for the dual-dynamo hypothesis has been found in the temporal analysis of stellar cycle data for Sun-like stars (Baliunas {\it et al.}, 1995; Ol\'ah {\it et al.}, 2009; Metcalfe {\it et al.}, 2013; Egeland {\it et al.}, 2015).
Alternatively, non-dynamo-based interpretations have also been proposed, involving hydrodynamic (HD) Rossby-type waves (Wolff, 1992; Lou, 2000; Sturrock {\it et al.}, 2015) and magnetic Rossby waves in the solar tachocline leading to the periodic emergence of magnetic flux at the solar surface due to magnetic buoyancy (Zaqarashvili {\it et al.}, 2010). 
Notwithstanding the above, intermediate-range periodicities are not continuously detected in solar data sets and it is also possible that the bulk of these variations is merely attributable to stochastic processes of magnetic-flux emergence through the photosphere and interaction, through magnetic reconnection, with existing coronal structures and plasma flows (see Wang and Sheeley (2003) for a discussion of this topic).
For this reason, the use of a specific technique that is able to distinguish pure oscillatory signals from colored noise at an appropriate significance level (say, $>99\,\%$) can be an important tool for determining the actual presence and evolution of these oscillations. 

The purpose of this article is to investigate the presence of QBOs in a set of intensity data of the green corona covering as long as six full solar cycles (18 through 23) from 1944 to 2008.
The green emission forbidden line of the solar corona (\fe, 530.3 nm) is the brightest coronal line in the optical range and is mainly detected in dense loops and loop clusters of the inner corona at a temperature of about $2 \times 10^6$ K.
Its intensity depends on the temperature and density of the coronal plasma, both parameters being modulated by the solar magnetic field. 
The emission from this line is thus a useful tracer of large coronal structures in both the quiet and the active corona. 
An advantage of the green-line emission is that it can be acquired on a daily basis almost simultaneously at the solar limb over all heliographic latitudes, thus allowing analysis of the spatio--temporal evolution of solar activity based on uniform data for the entire solar surface. 
Previous work has already established the presence of intermediate oscillations in the green corona through various techniques, such as the proper orthogonal decomposition (POD) analysis (Vecchio and Carbone, 2009), the empirical mode decomposition (EMD), and the Morlet wavelet transform (Deng {\it et al.}, 2005).
Here, we apply an alternative, advanced data-driven method, the multichannel singular spectrum analysis (MSSA) algorithm (see Ghil {\it et al.}, 2002 for a comprehensive review) that is specifically designed to empirically infer the characteristics of the space--time variations of complex systems and identify coherent space--time patterns within a given set of data.
The MSSA technique is particularly suitable to investigate the presence of intermediate-range oscillations in the large multivariate data set represented by the green-corona observations.
Unlike traditional spectrum analysis, where the basis functions are sinusoidal functions, MSSA has the advantage of being determined from estimates of the lagged cross-covariance where the basis functions are data-adaptive, empirical, and orthogonal.
If an oscillation is superimposed on colored noise with power around the frequency of the given oscillation, MSSA is able to distinguish between the part due to the oscillation and the part attributable to the noise.
By applying a Monte Carlo significance test, it is finally possible to establish whether the detected oscillations can be distinguished from colored noise with a specified confidence interval.

The remainder of the article is organized as follows: 
Section 2 presents the data set. 
Section 3 describes the detailed methodology of MSSA. 
In Section 4, we analyze the \fe\ data set with MSSA and identify spatio--temporal patterns of variability.
In Section 5 the results are discussed and conclusions are summarized in Section 6.

\section{Data}

Since 1939, systematic observations of the brightness of the coronal green line were carried by a network of several high-latitude coronal stations (Lomnick\'y St\'it, Sacramento Peak, Norikura, Kislovodsk, Pic du Midi, Wendelstein, Arosa, Kanzelh\"ohe) of which Lomnick\'y St\'it is the primary station since 1965.
The coronal intensities are recorded daily in tabular form, each set of measurements including a series of limb observations with a step of 5\dg\ in position angle, starting from the North Solar Pole counterclockwise.
Synoptic maps with a resolution of about 13\dg\ in longitude and 5\dg\ in latitude can be thus directly constructed from the observational data.
Coronal intensities are given in millionths of the intensity of the solar disk (coronal units) and converted to the photometric scale of the Lomnick\'y St\'it Station at a height of $40\arcsec$ (1.04 \rsun) above the solar limb (Rybansky {\it et al.}, 1994).
The full data set is available at the NOAA/NGDC ftp site ({\sf ftp://ftp.ngdc.noaa.gov/STP/SOLAR\_DATA/SOLAR\_CORONA/}).
The coronal green-line intensity data used in this work cover as long as six full solar cycles (18 through 23) from 1944 to 2008.
Figure 1 shows the latitude distribution of the coronal green emission, with the typical appearance of a butterfly diagram due to the migration of activity from the mid-latitudes toward the Solar Equator. 

From an observational point of view, in contrast to other more common tracers of solar activity such as sunspot numbers and areas, the \fe\ spectral line has the advantage that it has a latitude distribution much wider than the sunspot belt, so that periodicities as a function of latitude can be better studied with this line.
Moreover, while sunspot properties are mainly related to the cyclic emergence of magnetic-field flux tubes in active regions, the green corona, being widely distributed all over the Sun, implicitly contains  information on the global magnetic field of the Sun.
In fact, its intensity depends on the density and temperature of the plasma in the outer solar atmosphere, and both of these quantities are modulated by the local magnetic fields.

\begin{figure}
\centering
\includegraphics[width=11cm]{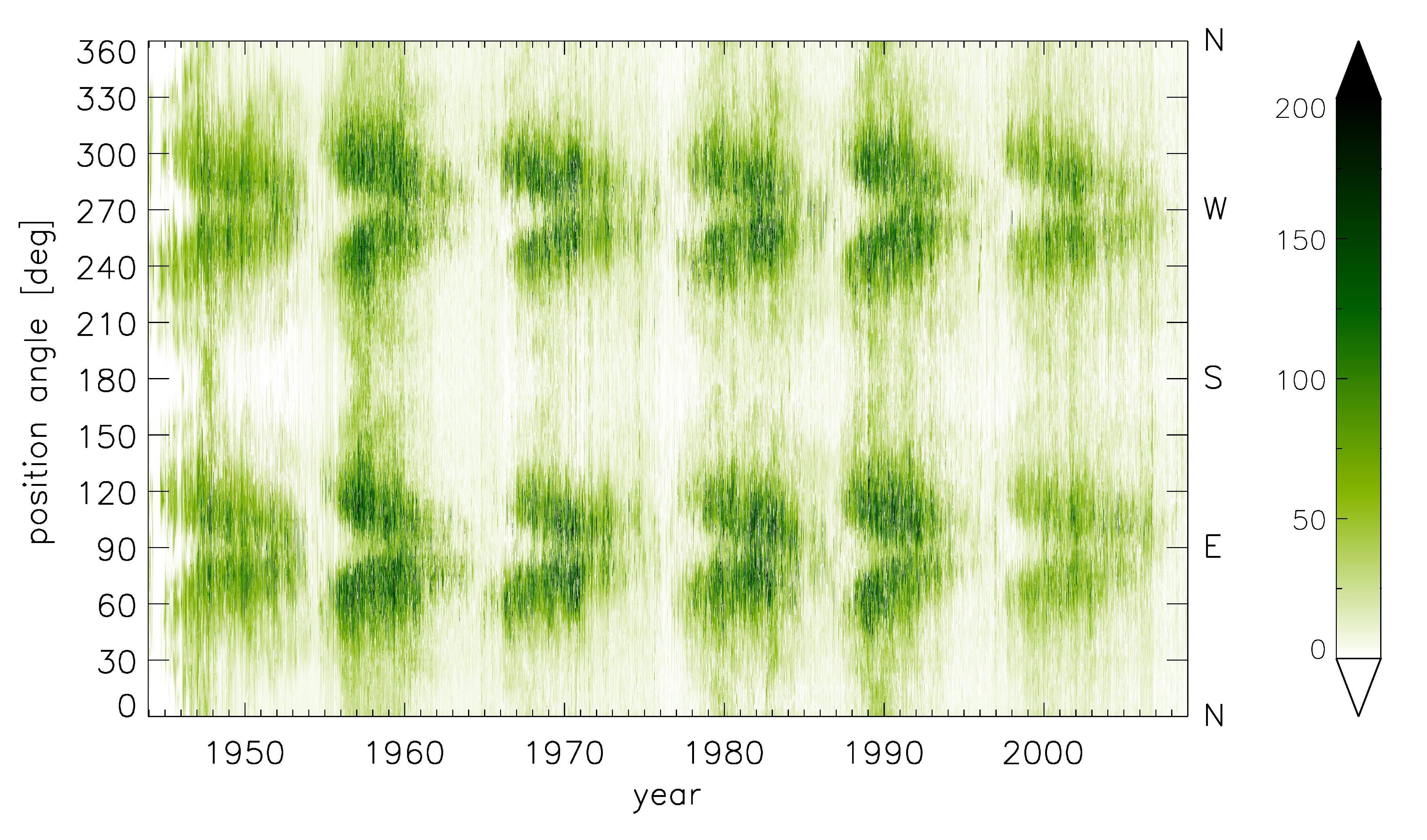}
\caption{The East- and West-limb daily intensity emission of the green coronal emission line \fe\ at 530.3 nm in the period $1944-2008$ as a function of time and position angle (counterclockwise from the North Pole at 0\dg). Coronal intensities are given in millionths of intensity of the solar disk (coronal units) and converted to the photometric scale of the Lomnick\'y St\'it Station at a height of $40\arcsec$ (1.04 \rsun) above the solar limb.}
\end{figure}

\section{Methodology}

The decomposition of a multiperiodic time series into separate oscillatory components traditionally relies on Fourier techniques that fit a linear combination of unmodulated sine and cosine functions.
Over the last few decades, an advanced data-adaptive technique called singular spectral analysis (SSA), as well as its multivariate extension (that is, MSSA), has been widely used in the identification of oscillations in climatic time series ({\it e.g.}, Taricco {\it et al.}, 2015a,b, 2016).
Although not commonly used in astrophysical contexts, this powerful technique has been recently applied with success by Mancuso and Raymond (2015) and Mancuso {\it et al.} (2016) for the analysis of Doppler-shift oscillations in the ultraviolet corona.
The major goal of SSA consists in the identification of pure oscillatory signals from noise in the analysis of nonlinear dynamics in univariate contexts: the original time-evolving signal is not simply decomposed into periodic sinusoidal functions as in Fourier-like techniques, but rather decomposed into data-adaptive waves which can be modulated both in amplitude and phase. 
Its multivariate extension, MSSA, has the added ability to identify coherent space--time patterns and thus extract common periodic signals, trends and noise from a multivariate data set. 
Moreover, MSSA is much more flexible than the standard methods of modeling that involve at least one of the restrictive assumptions of linearity, normality, and stationarity.

The theoretical framework of MSSA was proposed initially by Broomhead and King (1986) and later developed by Plaut and Vautard (1994) and Ghil {\it et al.} (2002).
This data-adaptive technique includes two stages: decomposition and reconstruction. 
In the first stage, MSSA decomposes a multivariate time series $X_l(t)$, with $t=1, \ldots, N$ representing time and $l=1, \ldots, L$ the individual time series (or channels), into an orthonormal, data-adaptive space--time structure whose elements represent eigenvectors of a grand lag-covariance matrix of size $LM \times LM$, where $M$ is the width of a sliding $M$-point window. 
Diagonalizing the above matrix results in a set of $LM$ eigenvectors $\bm {E}^k$, with $1 \leq k \leq LM$, called Space--Time Empirical Orthogonal Functions (ST-EOFs). Their associated Space--Time Principal Components (ST-PCs) $\bm {A}^k$, of time length $N' = N - M + 1$, are single-channel time series that are computed as
\begin{equation}
 A^k(t)  = \displaystyle\sum_{j=1}^{M}\sum_{l=1}^{L}  X_l(t+j-1) E^k_l(j),
\end{equation}
where $t$ varies from 1 to $N'$.
The $LM$ real eigenvalues $\lambda_k$, each associated with the $k$th eigenvector $\bm {E}^k$, equal the variance in the $\bm {A}^k$ direction.
The ST-PCs thus represent the different oscillatory modes extracted from the data set although, because of the lag window, they cannot be located into the same index space with the original time series. 
It is possible, however, to represent the same information in the original coordinate system by means of the so-called reconstructed components (RCs). 
Thus the $k$th RC at time $t$ for channel $l$ is given by \\

$R_l^k(t) =  {1 \over t}       \sum_{j=1}^{t}       A^k(t-j+1) E^k_l(j)   \quad  {\rm for~~~~}  1 \leq t \leq M-1 $  \\

$R_l^k(t) =  {1 \over M}     \sum_{j=1}^{M}      A^k(t-j+1) E^k_l(j)   \quad  {\rm for~~~~}  M \leq t \leq N-M+1$ ~~~~~~~~~~~~~~~~~~~~~~~~~~~~(2)\\

$R_l^k(t) =  {1 \over N-t+1} \sum_{j=t-N+M}^{M} A^k(t-j+1) E^k_l(j)   \quad  {\rm for~~~~}  N-M+2 \leq t \leq N $ \\

Each RC allows reconstruction of the dynamical behavior in $\bm {X}$ that belongs to $\bm {E}^k$.
By summing up all of the individual RCs, it is finally possible to recover the original time series, so that information is not lost in the decomposition and reconstruction process.

An important characteristic of the MSSA technique is that it may be used to identify modulated oscillations in the presence of colored noise. 
In the single-channel case, any oscillation detected through a window of width $M$ can be described completely in terms of only two vectors, a sine and cosine
with periods equal to the oscillation, provided that the period is less $M$ and that the time scales of amplitude- and phase-modulation are much greater than $M$ (Vautard and Ghil, 1989). 
If the variance of a series is dominated by such an oscillation, SSA will generate a pair of sinusoidal EOFs that will result in quadrature, that is, $\pi/2$ out of phase with each other and with similar amplitude. 
In a similar way, in the multichannel context, an oscillatory mode can be represented by MSSA as a pair of spatio--temporal patterns that are sinusoidal in time, have similar amplitude and are $\pi/2$ out of phase (Plaut and Vautard, 1994).
After MSSA, a test of statistical significance is needed to avoid spurious oscillations that may be due to non-oscillatory processes, such as first-order autoregressive AR(1) noise.
The oscillatory modes identified with MSSA can be tested against a red-noise null hypothesis through a Monte Carlo simulation (Allen and Robertson, 1996).
This null hypothesis is that the data have been generated by $L$ first-order, autoregressive independent processes ({\it i.e.} red noise).
The data set generated by the red-noise model is called the surrogate data set and it is subjected to MSSA in the same way as the original data set.
A large number of surrogates is generated to estimate the confidence limits for the MSSA result of the original data set. 
The null hypothesis can be rejected if the spectrum of the eigenvalues associated with the modes detected by MSSA is higher than that expected in the data generated by red-noise processes. 
As suggested by Groth and Ghil (2011), we also rely on a subsequent Varimax rotation of the ST-EOFs to improve the separability of distinct frequencies.

\section{Analysis and Results}

Some preprocessing of the original data sets is crucial to assess the statistical significance of the intermediate-range variability in the green corona using MSSA.
In fact, the \fe\ records, provided as daily averages, are dominated by a strong signal arising from the 27-day synodic period due to solar rotation and by its higher-order harmonics.
Since we are interested in periodicities $\gtrsim$ one year, we first applied a tapered Gaussian shaped filter with a FWHM of 54 days to the original time series.
The resulting series were then averaged at 27-day intervals, thus obtaining a total of $N = 880$ bins per channel. 
The application of the filter before the binning process is necessary to ensure that signal distortion due to aliasing from the large modes in the 27-day band is negligible. 
The resampling has also the advantage of reducing the total number of points in the time series, which greatly benefits the application of MSSA from a computational point of view. 
As usual in multivariate analysis, each time series has been non-dimensionalized by subtracting the mean and dividing by its standard deviation to avoid overweighting the grid points with higher variance.

\begin{figure}
\centering
\includegraphics[width=10cm]{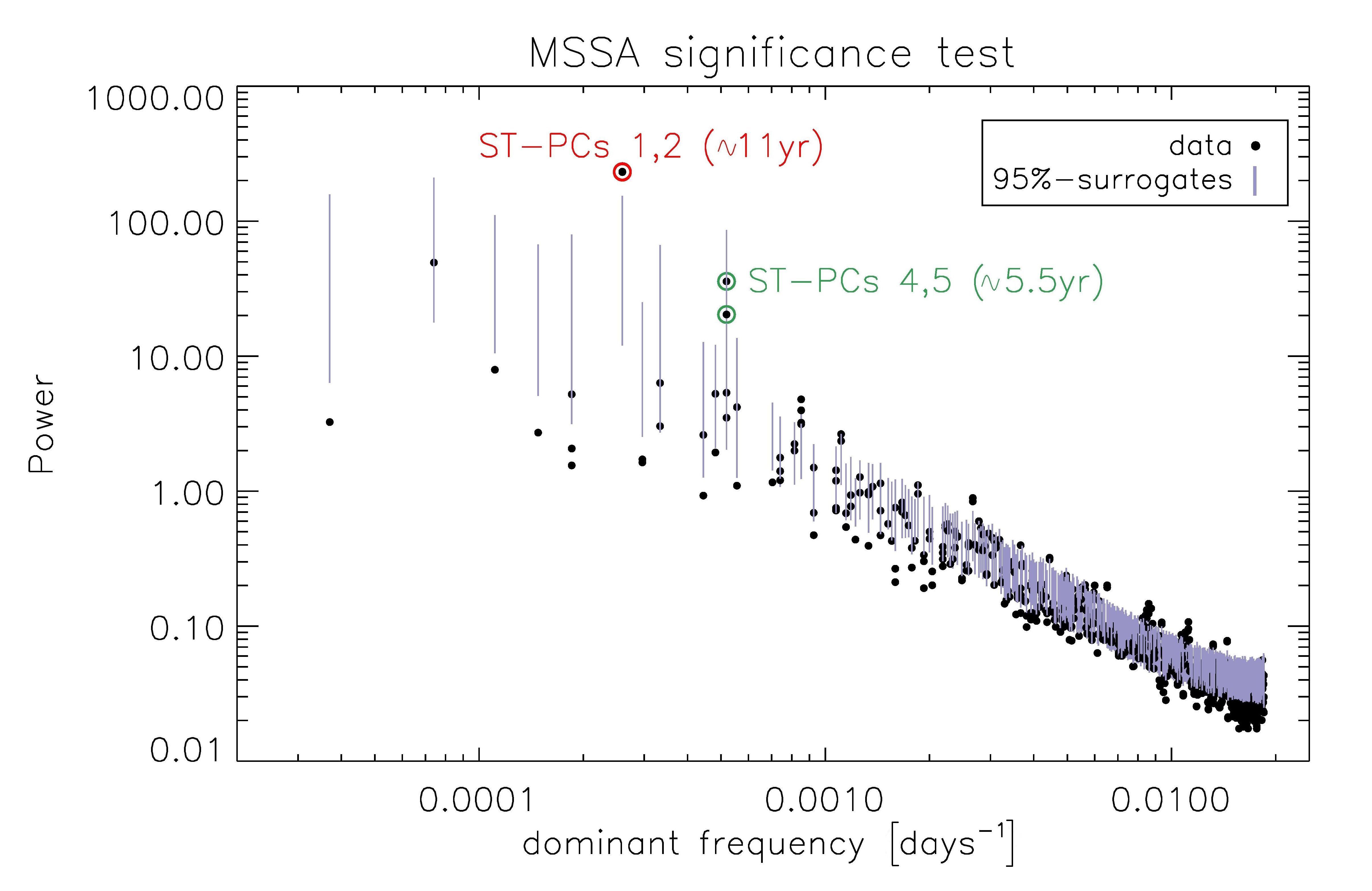}
\caption{
Monte Carlo significance test of \fe\ data for the whole Sun for the period 1944\,--\,2008, using all time series available as input channels over this interval. 
Shown are projections of the \fe\ data onto the data-adaptive basis, with a $\approx 11$-year window ($M = 146$). 
Filled circles show the data eigenvalues, plotted against the dominant frequency of the corresponding ST-PCs. 
The vertical bars give the 95\,\% confidence interval computed from 1000 realizations of a noise model consisting of $L$ independent AR(1) processes with the same variance and lag-1 autocorrelation as the input data channels. 
The dominant periods of the significant components are also indicated.}
\end{figure}

The only user-defined parameter in the application of the MSSA technique is the embedding dimension (or lag window) of length $M$, which is a flexible parameter to be judiciously chosen by the data analyst.
This choice often reflects a trade-off between optimal spectral resolution, temporal resolution, and noise reduction (see Ghil {\it et al.}, 2002).
Increasing the window length $M$ provides a more detailed spectral decomposition of the data set, although this choice is restricted by the condition $M<N/2$, where $N$ is the number of series points.
On the other hand, too large a window implies a poorer temporal resolution.
Moreover, $M$ cannot be set too small due to the fact that MSSA is not able to distinguish between different oscillations of period longer than $M$.
Given the above premises, a good compromise for the MSSA of the data set of coronal \fe\ intensities was found by using a window length $M=146$, roughly corresponding to the mean length of the dominant 11-year solar cycle.
In this way, we obtain ST-PCs of length $N' = N - M + 1 = 54.4$ years.
After MSSA, the Monte Carlo significance test of Allen and Robertson (1996) was applied in order to investigate which oscillating patterns contained more variance than would be expected if the data were generated by red noise.
The result of the Monte Carlo significance test, computed from 1000 realizations of a noise model consisting of $L$ independent AR(1) processes with the same variance and lag-1 autocorrelation as the input data channels, is shown for the data-adaptive basis in Figure 2, where both data and surrogates are projected onto the data ST-PCs.
As expected, the 11-year periodicity (ST-PCs 1--2) is found to be significant at the 99\,\% confidence level (c.l.). 
This periodicity dominates the temporal variability in the coronal \fe\ intensity data and explains about 60.8\,\% of the total variance.
The presence of the second harmonic at 5.5 years (ST-PCs 4--5, with 7.4\,\% of the total variance), which arises from the non-harmonic asymmetric form of the solar cycle (for which the rises tend to be slightly steeper than the falls) and does not correspond to a real additional periodicities, is also expected (see, {\it e.g.}, Mursula, Usoskin, and Zieger, 1997; Bloomfield, 2004).
The latter periodicity, not significant above the 99\,\% c.l., is only marginally significant above the 95\,\% c.l. 

\begin{figure*}
\centering
\includegraphics[width=7.5cm]{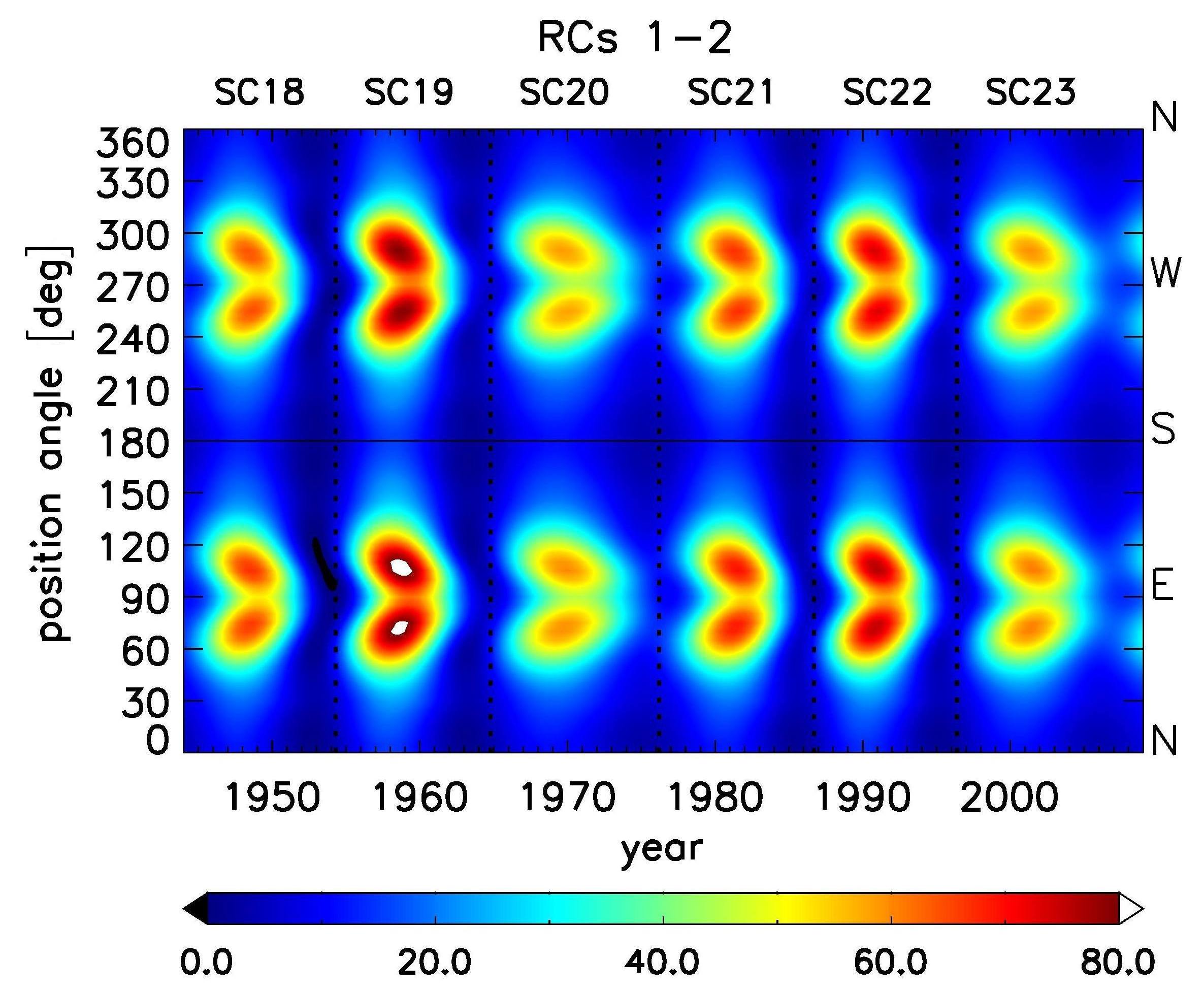}
\includegraphics[width=7.5cm]{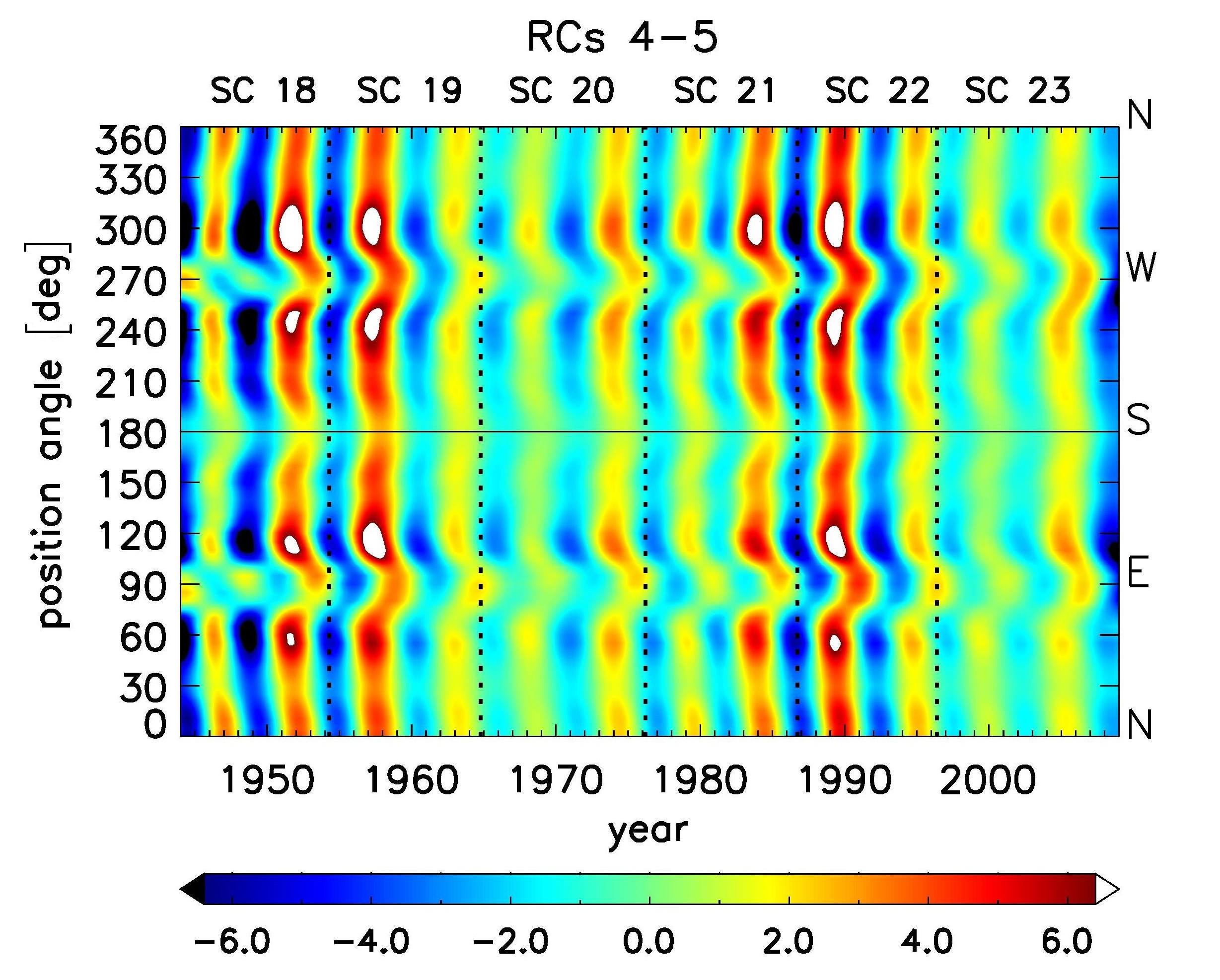}
\caption{
2D contour plots of the two pairs of reconstructed components RCs 1--2 ({\bf top panel}) and RCs 4--5 ({\bf bottom panel}). 
Values on the color bars of both plots are expressed in coronal units, as in Figure 1.}
\end{figure*}

Apart from the decomposition into data-adaptive modes, the most interesting aspect of MSSA is the detailed reconstruction of any subset of significant components (RCs) in the multivariate time series that can be obtained in much greater detail than other techniques, such as complex demodulation in classical Fourier analysis.
The resulting RCs are time series displaying the temporal evolution of each channel ({\it i.e.}, 5\dg\ latitude band) on the time scale of the corresponding MSSA mode.
Figure 3 shows the evolution in time of the coronal \fe\ emission intensity of the statistically significant RC pairs 1--2 and RCs 4--5 (representing, respectively, the 11-year and 5.5-year periodicities) reconstructed from the MSSA and expressed in coronal units, as in Figure 1.
The reconstructed diagram obtained from the RC pairs 1--2 (top panel of Figure 3) agrees quite well with the observed butterfly diagram as seen in Figure 1, albeit the high variability, complexity, and noise-related stochasticity displayed in the original data are now missing. 
The bottom panel of Figure 3 shows a similar reconstruction but obtained using the second pair of dominant components (RCs 4--5) corresponding to the 5.5-year periodicity. 
A slight East--West asymmetry in both RC pairs under study displayed in Figure 3 can be readily appreciated by visual inspection. 
This is expected and might be due only in part to some intrinsic calibration issue but largely to the fact that the corona is highly dynamic on scales of days or so because of the presence of localized active regions with associated flares and coronal mass ejections (CMEs). 

Several other modes appear to be significant at the 95\,\% c.l. in the MSSA spectrum shown in Figure 2, especially for periods $\lesssim$ one year.
We interpret the significant one-year component at frequency $\approx 0.0027$ days$^{-1}$ as mainly due to seasonal effects connected with the Earth's orbit around the Sun and the inclination of the solar rotation axis and/or the magnetic dipole axis. 
We point out that oscillations with periods below one year of order several months (often referred to as Rieger-type oscillations), although significant in the Monte Carlo test, will not be discussed in this article because they can only marginally be investigated with the MSSA technique. 
This is mainly due to the abovementioned restrictions in the choice of the length $M$ of the window and because of the 27-day averaging procedure applied in the preprocessing of the data that might produce residual aliases in our time series.
We also mention that the power in the range pertaining to Rieger-type oscillations drops to very low levels, with the result that it becomes increasingly difficult to distinguish possible signals from pure noise. 

\begin{figure}[t]
\centering
\includegraphics[width=8cm]{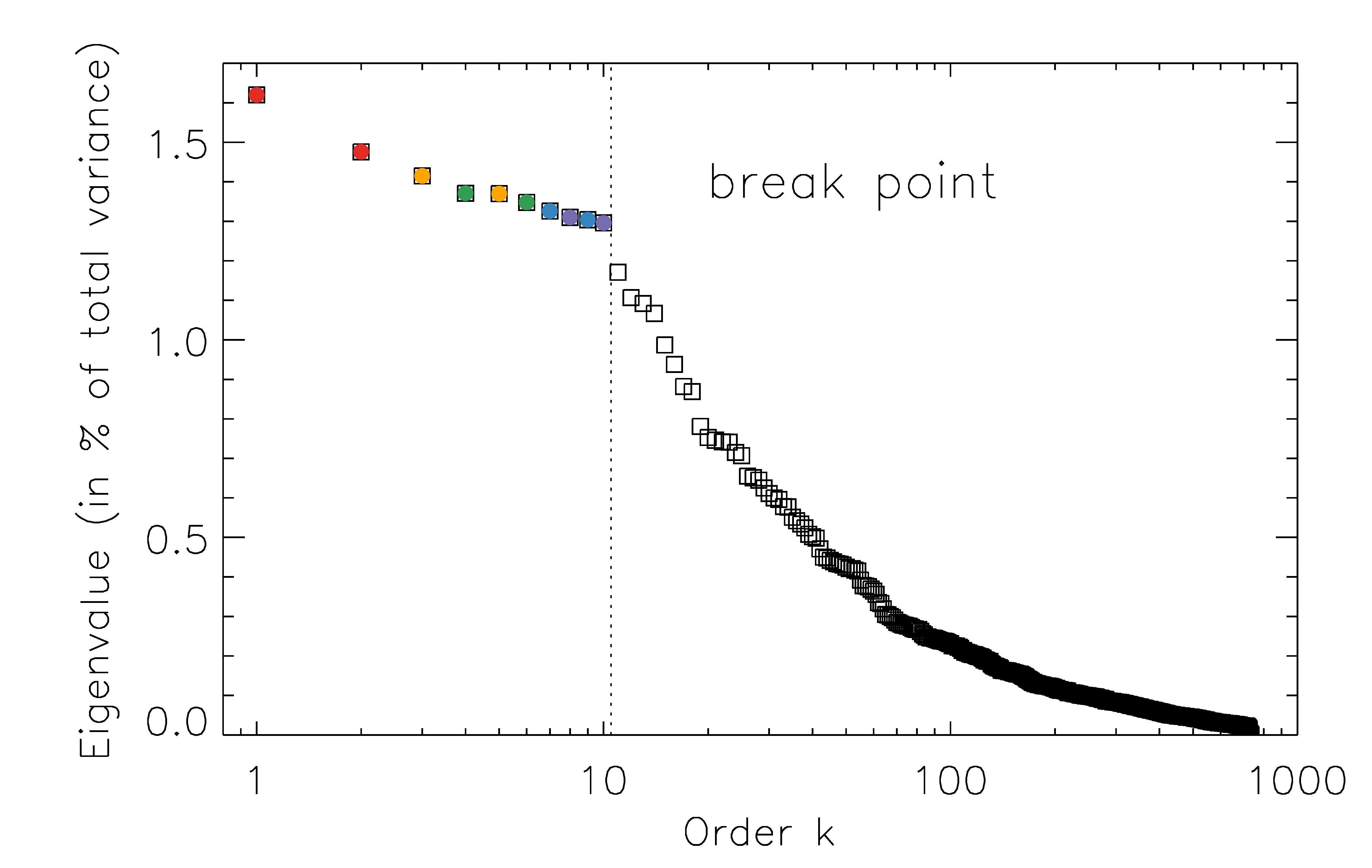}
\includegraphics[width=8cm]{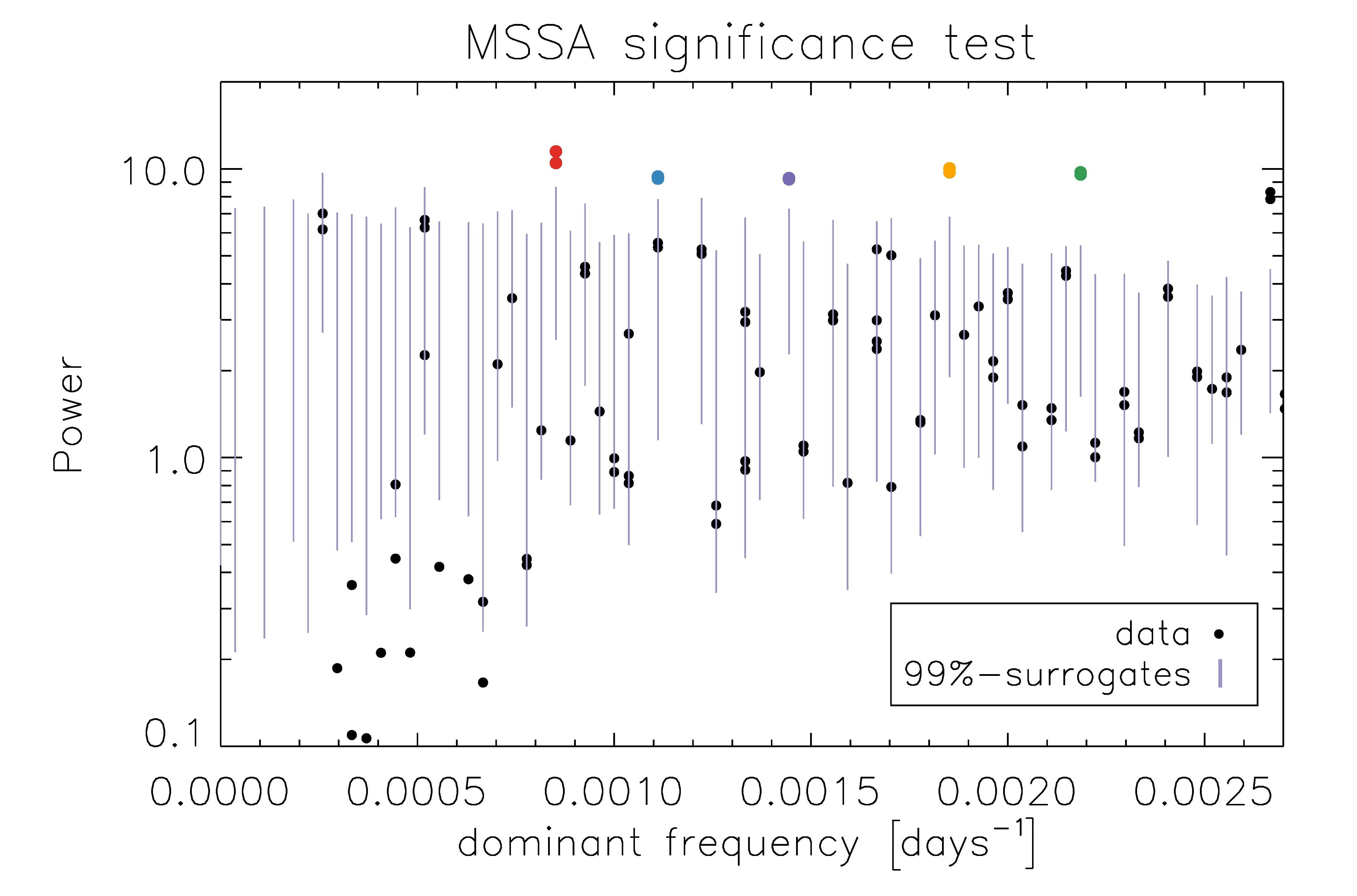}
\caption{
({\bf Top panel}) Rank-eigenspectrum obtained by applying the MSSA technique to the residual multivariate time series obtained after removing both the dominant 11-year signal of the solar cycle and its second harmonic.
The separation at the break-point between significant ($>99\,\%$ c.l.) periodic components and the low plateau of the noise components is shown by the vertical line.
({\bf Bottom panel}) Monte Carlo significance test. Shown are projections of the \fe\ data onto the data-adaptive basis, with a $\approx 11$-year window ($M = 146$). 
Closed circles show the data eigenvalues plotted against the dominant frequency of the corresponding ST-PCs. 
The vertical bars give the 99\,\% confidence interval computed from 1000 realizations of a noise model consisting of $L$ independent AR(1) processes with the same variance and lag-1 autocorrelation as the input data channels. 
The dominant periods of the significant components are also indicated with different colors.
}
\end{figure}

Since our aim is to investigate the intermediate oscillatory features of the solar-activity cycle corresponding to the QBOs and to discuss them in terms of spatio--temporal and dynamical complexity, we now proceed to determine the actual existence and significance of these modes in the multivariate data set, trying to distinguish them from the underlying noise, without losing relevant information. 
In order to better extract the periodicities under study, the data were further preprocessed by subtracting a smooth trend obtained by applying a tapered Gaussian shaped filter of width 4.5 years, from the original 27-day averaged time series. 
This process is applied to effectively remove both the dominant 11-year signal of the solar cycle and its second harmonic at 5.5 years, which, as we have seen, together explain about two-thirds of the total variance of the multivariate time series. 
Figure 4 shows the eigenvalue spectrum (or ``scree diagram") obtained by applying the MSSA technique to the residual multivariate time series with a window of length $M=146$.
In a scree diagram, the eigenvalues $\lambda_k$ are sorted in descending order so that the plot of eigenvalue $\lambda_k$ {\it vs.} $k$ represents a ranking of the relative variance contribution for each ST-PC.
Here we are mainly interested in the ST-PCs contributing a large fraction of the variance and representing oscillatory components of the signal, which stand out and can be distinguished from the underlying noise in the record.
In this case, signal-to-noise separation can be promptly obtained by merely inspecting the observed slope break in the scree diagram of eigenvalues. 
The diagram in the top panel of Figure 4 clearly shows a group of ten eigenvalues (filled colored boxes) with slowly decreasing values of variance, followed by a steeper slope of the remaining additional eigenvalues (empty boxes).
The latter are well separated from the first ten eigenvalues, forming a mildly sloping and flattening out ``tail" of the MSSA spectrum.
In this case, the cutoff for ``significant" components is simply determined to be the value of $k$ where the change in slope of the ranked eigenvalues is observed.
Having established the ``noise floor" from the break in the slope of the scree diagram, the remaining analysis will be focused on the first ten components, up to the point where the slope of  $\lambda_k$ {\it vs.} $k$ changes (that is, up to $k=10$) and the variance contribution is smaller, which thus represent the discarded noise.
The leading ten ST-PC time series and the corresponding normalized periodograms are displayed in Figure 5, ordered according to the explained variance.

\begin{figure}
\centering
\includegraphics[width=11cm]{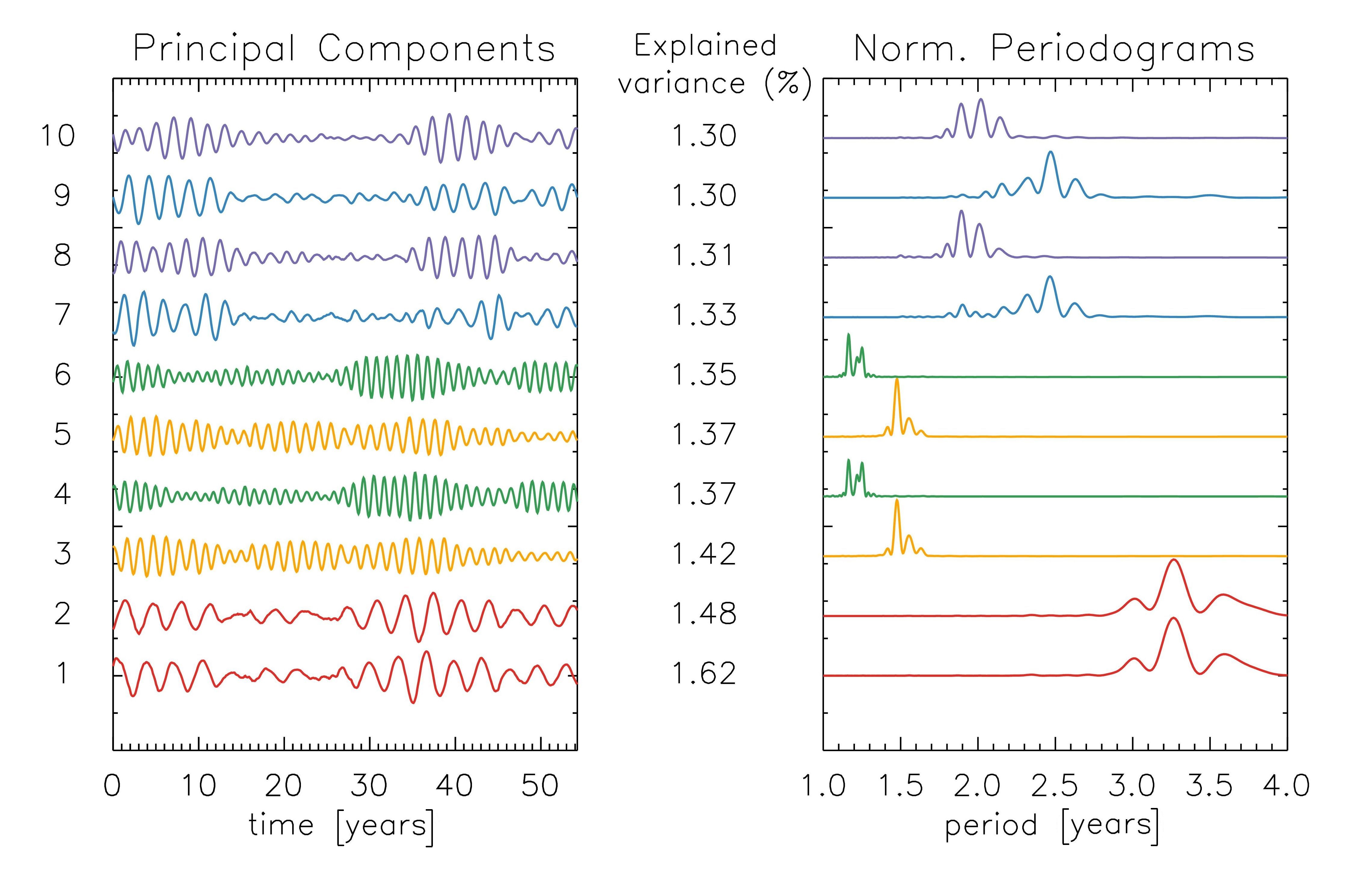}
\caption{
{\bf (Right panel)} Principal components of the ten dominant and significant ($>99\,\%$) oscillatory modes for the MSSA obtained by filtering out all periods above 4.5 years and corresponding normalized periodograms ({\bf left panel}). 
The components are ordered according to the explained variance [\%]. 
The time span on the horizontal axis in the left panel corresponds to the length expressed in years of the ST-PCs, that is, $N' = N - M + 1$; see text for details.}
\end{figure}

\begin{table}[b]
\caption{Significant ($> 99\,\%$ c.l.) oscillating ST-PC pairs.}  
\begin{tabular}{ccc} 
\hline
ST-PC pair          & Signif. period [years] & \% of total variance \\ \hline 
1 and 2                    & 3.2              &     3.10          \\
3 and 5                    &  1.5             &     2.78           \\ 
4 and 6                   &  1.3              &     2.72           \\ 
7 and 9                  &  2.5               &     2.63           \\ 
8 and 10                  &  1.9             &     2.61           \\  
\hline 
\end{tabular} 
\end{table}

As shown in Figure 5, the ten ST-PCs actually form by themselves five distinct pairs of oscillating components (ST-PCs 1--2, 3 and 5, 4 and 6, 7 and 9, 8 and 10) that are alike, in quadrature (see left panel of Figure 5) and share a very similar frequency in the QBO range (see right panel of Figure 5): these two facts together support the oscillatory nature of the above modes (Plaut and Vautard, 1994).
In Table 1, the five oscillating ST-PC pairs are listed together with their period $P$ [years] identified by applying the multitaper spectral analysis method (MTM; Thomson, 1982; Mann and Lees, 1996). 
Corresponding variances are also shown in \% of total multivariate time series' variance.
The first ten eigenvalues, with periods in the range 1.3 to 3.2 years capture as much as 13.84\,\% of the total variance.
The MSSA results thus confirm the presence of distinct QBO modes in the data.
As for their significance, the Monte Carlo test of Allen and Robertson (1996) for MSSA was also applied in this case in order to investigate whether the oscillating patterns found in the quasi-biennial range contained more variance than would be expected if the data were generated by red noise.
We verified that the ten leading MSSA eigenmodes identified in the quasi-biennial range stood out strongly in the spectra of eigenvalues, forming oscillatory pairs statistically significant against red noise at 99\,\% c.l. (see bottom panel of Figure 4).

Even more useful than examining the ST-PCs individually (or in pairs) is the process of reconstructing the multivariate space--time series associated with each pair (or the entire subset) as explained in the previous section. 
We used the RCs for visualization of the spatio--temporal patterns related to the significant ST-PCs that characterize both the spatial and the temporal structure of the modes of variability in the QBO range of interest.
In Figure 6, we show 2D contour plots of the reconstructed signals associated with the five significant ST-PC pairs listed in Table 1.
By visual inspection, we notice a good symmetry between the results obtained from the East and West limbs, although a few differences here and there stand out. 
This discrepancy is mainly attributable to the already mentioned dynamic nature of the corona and the fact that the standard deviation of the intensity of the coronal emission is somewhat stronger in the East limb with respect to the West limb.
A major result that can be derived from the analysis of the above contour plots is the presence of a distinct asymmetry in the distribution of the QBOs among the northern and southern hemisphere of the Sun. 
This latter could be related in turn to an intrinsic latitudinal asymmetry in the mechanism that produces such higher frequency modes.  
This is not the first report of North--South asymmetry in QBOs: this puzzling behavior has been already reported in the literature (see, {\it e.g.}, Badalyan and Obridko, 2004, 2011; Knaack, Stenflo, and Berdyugina, 2004; Forg\'acs-Dajka and Borkovits, 2007; Badalyan, Obridko, and S\'ykora, 2008).
Badalyan and Obridko (2011) even suggested that solar activity may be actually generated independently in the two hemispheres. 

\begin{figure}
\centering
\includegraphics[width=12cm]{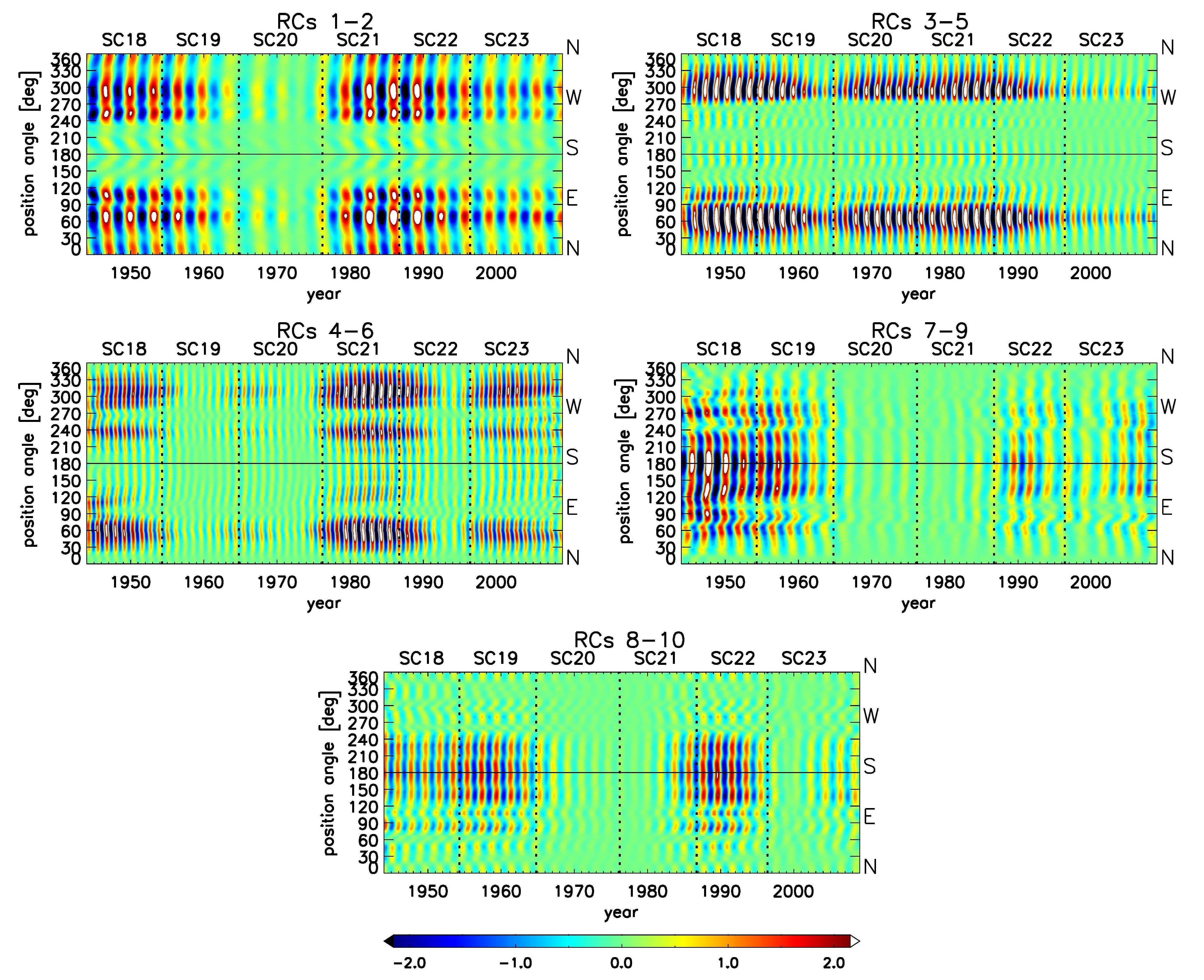}
\caption{
2D plots of the QBO oscillations represented by the five pairs of RCs listed in Table 1. }
\end{figure}

In the first pair of reconstructed components, RCs 1--2, with a nominal period of 3.2 years as identified by the MTM technique (see Table 1) but actually corresponding to periods in the range between about 2.9 and 4.0 years (see right panel of Figure 5), the QBO signal appears mainly concentrated in a range of latitudes around the solar equator, definitely corresponding to the sunspot belt.
This lower-frequency QBO signal is, however, enhanced in the polar region of the northern hemisphere (and therefore also related to the global magnetic field of the Sun), fading in Solar Cycle 19 and almost disappearing during Solar Cycle 20. 
The signal is also quite weak during the previous Solar Cycle 23. 
Low-frequency QBOs corresponding to periods in the range between about 2.9 and 4.0 years have not been frequently reported in the literature (see Table 1 in Bazilevskaya {\it et al.}, 2014) with the notable exception of Vecchio and Carbone (2009) who used a different technique to investigate the same set of data analyzed in this work. 

The second most powerful significant pair of reconstructed components, RCs 3 and 5, represent QBOs with a nominal period of 1.5 years (as identified by the MTM technique) but actually corresponding to QBOs with periods in the range between about 1.4 and 1.7 years (see right panel of Figure 5).
Remarkably, the QBO signal represented by this RC pair is mainly enhanced at mid-latitudes in the northern hemisphere but virtually absent in the southern hemisphere, almost disappearing during Solar Cycle 23.
The latitudinal distribution of the highest-frequency QBO signal reconstructed in the RC pair 4 and 6, corresponding to QBOs with periods between about 1.2 and 1.4 years, is again mainly (but not exclusively) evident at mid-latitudes in the northern hemisphere, being particularly enhanced during Solar Cycles 18, 21, and 23.
Quasi-periodicities in the range covered by the RC pair 3 and 5 and the RC pair 4 and 6 (1.2 to 1.7 years) have been already detected in the literature.
Helioseismic probing of the solar interior has shown that the rotation rate of the Sun near the base of the convective zone changes with a period of roughly 1.3 years (Howe et al., 2000).
Wavelet analysis of sunspot areas and sunspot numbers shows that a periodicity of 1.3 years is also present in these records of solar activity (Krivova and Solanki, 2002).
This periodicity has even been detected in variations of the interplanetary magnetic field and geomagnetic activity (Paularena, Szabo, and Richardson, 1995; Szabo, Lepping, and King, 1995; Lockwood, 2001) and in the solar-wind speed (Richardson et al., 1994).
As well as with the 1.3 years periodicity, a distinct $\approx 1.7$ quasi-periodic variation in the {\it Voyager} 1 and 2 data for $>70$ MeV protons has been detected in agreement with similar variations observed by neutron monitors ({\it e.g.}, Kato {\it et al.}, 2003).

As for the remaining two RC pairs, RC pair 7 and 9 and RC pair 8 and 10, which represent QBOs with nominal periods 2.5 and 1.9 years, respectively, as identified by the MTM technique (but altogether corresponding to intermediate periods between about 1.7 and 2.7 years), the QBO signal is mainly concentrated in the southern hemisphere, above the active-region belt of the Sun, thus suggesting its relation to the global magnetic field.
QBOs with periods in this range (1.7 to 2.7 years) have also been reported in the literature ({\it e.g.}, Obridko and Shelting, 2001; Ryb\'ak, Antalov\'a, and Storini, 2001; Shirai, 2004; Cadavid {\it et al.}, 2005; Vecchio and Carbone, 2008; Fletcher {\it et al.}, 2010; Vecchio {\it et al.}, 2010; Laurenza {\it et al.}, 2012).
The signal is particularly strong during Solar Cycles 18 and 19, almost disappearing in the following two solar cycles, just to reappear in Solar Cycle 22 and (weakly) in Solar Cycle 23.
By visual inspection of Figure 6, the signal contained in these two latter RC pairs seems to be (rather loosely) a complementary counterpart (albeit at a higher frequency) of the strong signal revealed in the northern hemisphere by RCs 1--2.
That is, the temporal evolution of the QBOs with intermediate periods between about 1.7 and 2.7 years is similar to the ones with longer periods between above about three years, maybe hinting at a similar mechanism exciting both of these lower frequency groups of modes but with an asymmetric North--South spatial distribution.  
Although the mechanism producing the two sets (long and intermediate period) QBOs could be similar, it is not clear, however, for what physical reason QBOs with longer period should be preferably excited in the northern hemisphere while QBOs with intermediate period  should be preferably excited in the southern hemisphere.

\begin{figure}
\centering
\includegraphics[width=12cm]{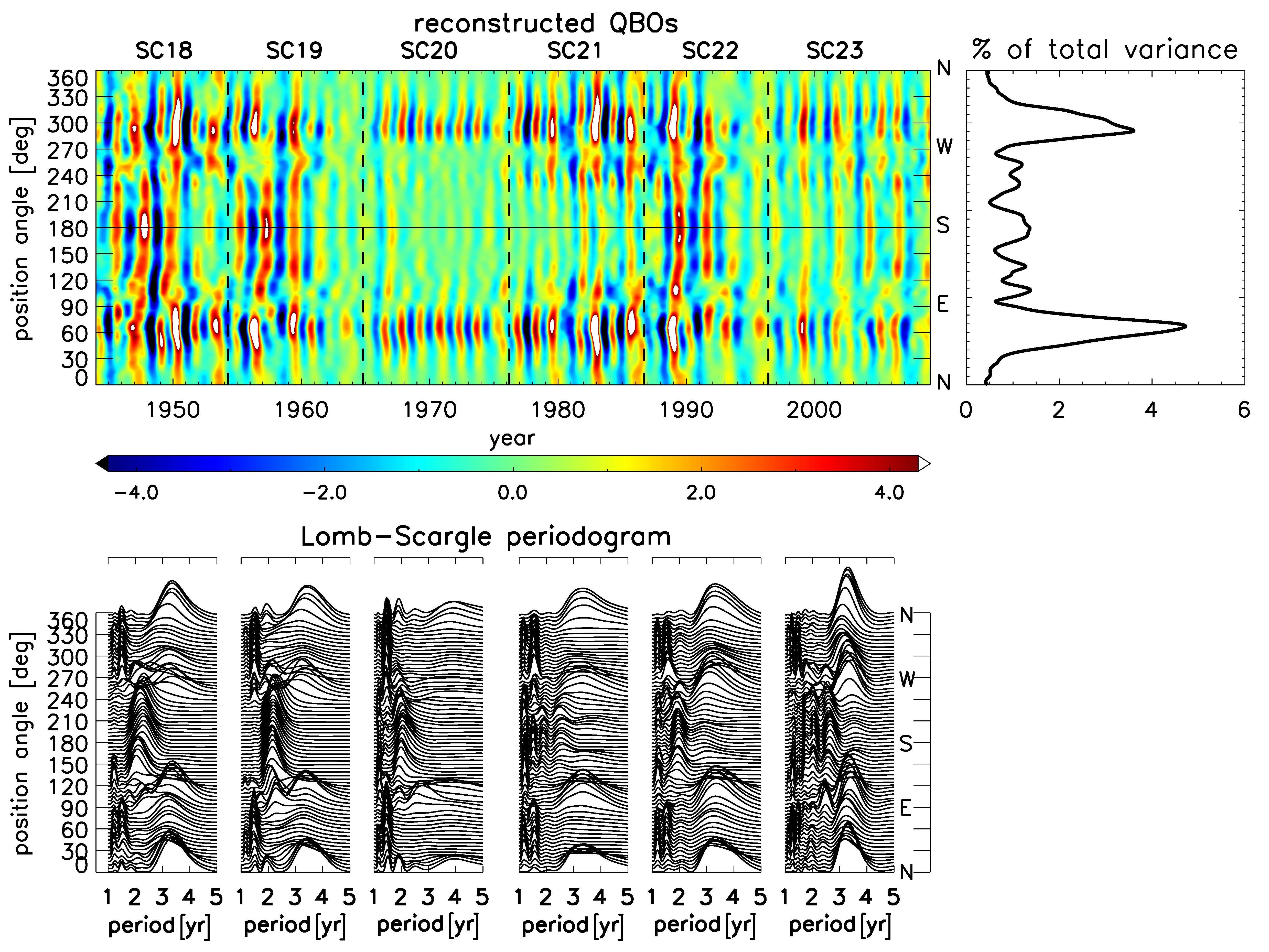}
\caption{
{\bf (Top panel)}: 2D plot of the signal associated with the subset of ten RCs pairs listed in Table 1 and percentage of total variance as a function of latitude.
{\bf (Bottom panel)}: Normalized periodograms of the reconstructed multivariate time series obtained by combining all pairs of RCs listed in Table 1 for each solar cycle from 18 to 23. 
}
\end{figure}

Figure 7 yields a global view of the spatio--temporal distribution of QBOs as obtained in this work.
The 2D plot in the top panel was obtained by summing up all ten reconstructed components listed in Table 1.
The North--South asymmetry in the distribution of periods is evident in the plot showing the percentage of total variance as a function of latitude, which is clearly dominated by the power associated with the active-region belt in the northern hemisphere.
The plot at the bottom of Figure 7 shows, for each individual cycle, periodograms of the reconstructed time series, yielding a visual perspective of the spatio--temporal distribution of periodicities in the global QBO signal. 

Due to their distribution over all latitudes and their asymmetric behavior, the periodicities in the quasi-biennial range cannot be totally ascribed to the evolution and decay of active regions, thus advocating alternative mechanisms for their excitation.

\section{Discussion}

The intensity of the green corona depends on the density and temperature of the plasma in the outer solar atmosphere, both of these quantities being modulated by the coronal magnetic field.
The green corona thus contains information on the global magnetic field of the Sun.
From an analysis of synoptic magnetograms for the zonally averaged radial magnetic-field component at the solar surface obtained using data measured at the Mt. Wilson Observatory, Ulrich and Tran (2013) found a well-defined cyclic signal at low- to mid-latitudes in the quasi-biennial range that is clearly associated with the process that reverses the global magnetic dipole during the solar cycle.
The QBO oscillations detected in this work may thus represent a manifestation of this intermittent magnetic-flux migration from low latitudes in the poleward direction, as also suggested by Vecchio {\it et al.} (2012) from the analysis of NSO/Kitt Peak magnetic synoptic maps covering the period 1976\,--\,2003.
In support of this interpretation, we remark that the quasi-biennial periodicity (for periods $\gtrsim 1.7$ years) is well detected in the polar data, where the contribution to the solar cycle is mainly due to the large-scale solar magnetic field (see Figure 7).
This indicates that at least part of the QBO signal could be described with a mechanism that is different from the established dynamo process generating the emergence of active regions.
Multiple dynamo cycles have been inferred theorethically in some recent magnetohydrodynamic (MHD) simulations of solar convection (Simard, Charbonneau, and Bouchat, 2013; K{\"a}pyl{\"a} {\it et al.}, 2016) and actually deduced from the temporal analysis of stellar-cycle data for Sun-like stars (Baliunas {\it et al.}, 1995; Ol\'ah {\it et al.}, 2009; Metcalfe {\it et al.}, 2013; Egeland {\it et al.}, 2015).
It is thus plausible that the origin of the observed QBOs could be related to the presence of two different dynamo processes acting in the deep and near-surface layers of the convective zone that are responsible, respectively, for the sunspot cycle and the shorter variations corresponding to the observed QBO modes (see, {\it e.g.}, Benevolenskaya, 1998; Fletcher {\it et al.}, 2010; Obridko and Badalyan, 2014; Beaudoin {\it et al.}, 2016).
Our results are in agreement with the ones obtained by Vecchio and Carbone (2009) that, besides the main 11-years cycle, detected a 
well-defined high-frequency quasi-biennial component, mainly on the polar regions of the Sun, in the green coronal emission line recorded from 1939 to 2005. 
In their work, made by using proper orthogonal decomposition (POD) (Holmes, Lumley, and Berkooz, 1998) and wavelet analysis, Vecchio and Carbone (2009) interpreted the short-term component as likely originated from a stochastic superposition of different oscillators, probably generated by a turbulence-like phenomenon in a narrow band of frequencies.
We point out that the specific technique used in this work, being tailored to identify pure oscillatory signals from noise in the multivariate analysis of nonlinear dynamical systems at an appropriate significance level, is expected to better characterize the spatio--temporal evolution of these high-frequency modes, properly distinguishing pure oscillatory signals from colored noise, thus avoiding unwanted spurious signal contaminations.
As a result, the application of the MSSA technique to the green coronal emission line data has allowed a more reliable representation of the spatio--temporal evolution and North--South asymmetry (both in amplitude and period length) of the individual components of these periodicities (see Figure 6) together with a clearer global view of the spatio--temporal distribution of the QBOs in toto (see Figure 7).

As a cautionary note, we emphasize that some of the significant QBO modes found in this work may actually correspond to higher harmonics of the 11-year solar cycle.
Non-sinusoidal signals are known to show multiple peaks in the power spectrum at harmonic frequencies of the true periods.
In fact, considering that the average period of the six solar cycles analyzed in this work is 10.8 years, we do expect, at least in principle, the presence of higher harmonics with periods of about 5.4, 3.6, 2.7, 2.2, 1.8, 1.5, 1.3, ..., years.
These higher harmonics would not obviously correspond to independent additional periodicities but would arise from the asymmetric form or higher-order moments of the 11-year solar cycle.
Mursula, Usoskin, and Zieger (1997) used a simple nonlinear model to demonstrate explicitly how the asymmetric form of the sunspot cycle can produce the spectral power at the second harmonic of the fundamental 11-year periodicity and progressively decreasing peaks corresponding to higher harmonics. 
Looking at the results displayed in Table 1, it is clear that some (if not all) significant ST-PC pairs found in the MSSA could loosely correspond to these harmonics, allowing for the fact that we are not dealing with a model solar cycle but with real (and noisy) data.
As such, the high power detected in correspondence to the active region belt for the QBOs with longer periods (RCs 1--2 in Figure 6, corresponding to QBOs with periods in the range between about 2.9 and 4.0 years) could actually be related to the third harmonic of the solar cycle.
Disentangling the two contributions (that is, independent modes {\it vs.} higher harmonics) is, however, not possible with the present technique also considering the data-adaptive nature of the extracted modes that are generally spread (not being strictly sinusoidal due to their amplitude and frequency modulation) over a wider range of frequencies (see Figure 5).
Thus, the hypothesis that the QBOs as detected in the MSSA spectrum (and, even more so, in spectra obtained with Fourier-based techniques by other authors) could actually be the manifestation of higher harmonics (or contaminated by higher harmonics) of the fundamental 11-year period cannot be completely refuted.

In synthesis, we cannot exclude the possibility that the outcome of our multivariate data-adaptive analysis could represent a mixture of periodicities of both independent QBO modes arising from some physical mechanism in the interior of the Sun (such as the ones involving a double-dynamo hypothesis or non-dynamo interpretations based on Rossby-type waves), stochastic processes of magnetic-flux emergence through the photosphere as originally suggested by Wang and Sheeley (2003) and/or higher harmonics (up to the eighth or so) arising from the asymmetric form or higher-order moments of the 11-year signal due to the solar cycle (see also Polygiannakis, Preka-Papadema, and Moussas, 2003 and Kane, 2005 for further discussion of this topic).

\section{Conclusions}

We analyzed the spatio--temporal dynamics of the \fe\ green coronal emission line at 530.3 nm in the time interval from 1944 to 2008 to investigate the intermediate-range periodicities by means of a data-adaptive, multivariate technique called multichannel singular spectrum analysis (MSSA). 
Our analysis revealed the presence of significant oscillatory modes, with periods in a range from about one to four years, which are consistent with the so-called quasi-biennial oscillations (QBOs) that have been detected by several authors using different data sets and analysis methods.
QBOs of the coronal \fe\ emission intensity are present in each of the six cycles considered in this study, although not continuously (see Figure 7).
A clear North--South asymmetry is detected both in their intensity and period distribution, with a net predominance of spectral power in the active region belt of the northern hemisphere. 
Moreover, the amplitude of variations in the QBO range changes significantly with time between different cycles. 
Apart from their intermittent nature, being more or less excited depending on the individual solar cycle under study, QBOs have been found to have a frequency-dependent, uneven distribution in heliographic latitude.
In particular, higher-frequency oscillatory modes, with periods between about 1.2 and 1.7 years, are mainly (but not exclusively) concentrated in the northern hemisphere.
Intermediate oscillatory modes with periods between about 1.7 and 2.7 years are almost exclusively excited in the southern pole and are present in four of the six solar cycles examined in this study (but quite faint in the previous Solar Cycle 23), with the exception of Solar Cycles 20 and 21. 
Finally, lower-frequency oscillatory modes, with periods between about 2.7 and 4.0 years, are present in both hemispheres within the magnetically active streamer belt and enhanced in the northern hemisphere, especially at the Pole.
In summary, while the QBOs with longer and intermediate periods are particularly powerful around the polar regions, and therefore  related to the global magnetic field, the ones with shorter periods appear to be mainly generated at mid-latitudes, in correspondence to the emergence of active regions. 
This indicates that the group of QBO components with periods $\gtrsim 1.7$ years must be related to different physical mechanisms than the process generating the emergence of active regions, thus supporting the double-dynamo hypothesis or non-dynamo interpretations based on Rossby-type waves.

Our findings indicate that the North--South asymmetry manifested in the uneven latitudinal distribution of QBOs is a fundamental, albeit puzzling, characteristic of solar activity. 
Our results can thus provide more constraints on dynamo models put forward by theoreticians to describe the different components of the solar cycle.
In conclusion, we have introduced an alternative approach to extracting and describing the evolution of quasi-biennial oscillations from coronal \fe\ intensity time series, thus showing that MSSA is a viable and complementary tool for exploring the spatio--temporal behavior of intermediate oscillations from multivariate coronal time series.

\bibliographystyle{spr-mp-sola}

\begin{thebibliography}{}
\bibitem[\protect\citeauthoryear{{Berger}}{2003}]{Berger03b}Allen, M. R., Robertson, A. W.: 1996, \textit{Clim. Dyn.} \textbf{12}, 775. \adsurl{1996ClDy...12..775A}, \doiurl{10.1007/s003820050142}. 
\bibitem[\protect\citeauthoryear{{Berger}}{2003}]{Berger03b}Badalyan, O. G., Obridko, V. N.: 2004, \textit{Astron. Rep.} \textbf{48}, 678. \adsurl{2004ARep...48..678B}, \doiurl{10.1134/1.1787070}. 
\bibitem[\protect\citeauthoryear{{Berger}}{2003}]{Berger03b}Badalyan, O. G., Obridko, V. N.: 2011, \textit{New Astron.} \textbf{16}, 357. \adsurl{2011NewA...16..357B}, \doiurl{10.1016/j.newast.2011.01.005}. 
\bibitem[\protect\citeauthoryear{{Berger}}{2003}]{Berger03b}Badalyan, O. G., Obridko, V. N., S\'ykora, J.: 2008, \solphys{} \textbf{247}, 379. \adsurl{2008SoPh..247..379B}, \doiurl{10.1007/s11207-008-9120-0}.
\bibitem[\protect\citeauthoryear{{Berger}}{2003}]{Berger03b}Baliunas, S. L., Donahue, R. A., Soon, W. H., Horne, J. H., Frazer, J., Woodard-Eklund, L., {\it et al.}: 1995, \apj{} \textbf{438}, 269. \adsurl{1995ApJ...438..269B}, \doiurl{10.1086/175072}.
\bibitem[\protect\citeauthoryear{{Berger}}{2003}]{Berger03b}Bazilevskaya, G., Broomhall, A. -M., Elsworth, Y., Nakariakov, V. M.: 2014, \ssr{} \textbf{186}, 359. \adsurl{2014SSRv..186..359B}, \doiurl{10.1007/s11214-014-0068-0}.
\bibitem[\protect\citeauthoryear{{Berger}}{2003}]{Berger03b}Beaudoin, P., Simard, C., Cossette, J. -F., Charbonneau, P.: 2016, \apj{} \textbf{826}, 138. \adsurl{2016ApJ...826..138B}, \doiurl{10.3847/0004-637X/826/2/138}.
\bibitem[\protect\citeauthoryear{{Berger}}{2003}]{Berger03b}Benevolenskaya, E. E.: 1998, \apjl{} \textbf{509}, 49. \adsurl{1998ApJ...509L..49B}, \doiurl{10.1086/311755}.
\bibitem[\protect\citeauthoryear{{Berger}}{2003}]{Berger03b}Bloomfield, P.: 2004, \textit{Wiley Series in Probability and Mathematical Statistics}, Wiley and Sons, New York. 
\bibitem[\protect\citeauthoryear{{Berger}}{2003}]{Berger03b}Broomhead, D. S., King, G. P.: 1986, \textit{Physica D Nonlinear Phenomena} \textbf{20}, 217. \adsurl{1986PhyD...20..217B}, \doiurl{10.1016/0167-2789(86)90031-X}.
\bibitem[\protect\citeauthoryear{{Berger}}{2003}]{Berger03b}Cadavid, A. C., Lawrence, J. K., McDonald, D. P., Ruzmaikin, A.: 2005, \solphys{} \textbf{226}, 359. \adsurl{2005SoPh..226..359C}, \doiurl{10.1007/s11207-005-8187-0}.
\bibitem[\protect\citeauthoryear{{Berger}}{2003}]{Berger03b}Deng, L. H., Li, B., Xiang, Y. Y., Dun, G. T.: 2015, \jastp{} \textbf{122}, 18. \adsurl{2015JASTP.122...18D}, \doiurl{10.1016/j.jastp.2014.10.016}.
\bibitem[\protect\citeauthoryear{{Berger}}{2003}]{Berger03b}Egeland, R., Metcalfe, T. S., Hall, J. C., Henry, G. W.: 2015, \apj{} \textbf{812}, 12. \adsurl{2015ApJ...812...12E}, \doiurl{10.1088/0004-637X/812/1/12}.
\bibitem[\protect\citeauthoryear{{Berger}}{2003}]{Berger03b}Fletcher, S. T., Broomhall, A -M., Salabert, D., Basu, S., Chaplin, W. J., Elsworth, Y., Garcia, R. A., New, R.: 2010, \apjl{} \textbf{718}, L19. \adsurl{2010ApJ...718L..19F}, \doiurl{10.1088/2041-8205/718/1/L19}.
\bibitem[\protect\citeauthoryear{{Berger}}{2003}]{Berger03b}Forg{\'a}cs-Dajka, E., Borkovits, T.: 2007, \mnras{} \textbf{374}, 282. \adsurl{2007MNRAS.374..282F}, \doiurl{10.1111/j.1365-2966.2006.11167.x}.
\bibitem[\protect\citeauthoryear{{Berger}}{2003}]{Berger03b}Ghil, M., Allen, M. R., Dettinger, M. D., Ide, K., Kondrashov, D., Mann, M. E., {\it et al.}: 2002. \textit{Rev. Geophys.} \textbf{40}, 1003. \adsurl{2002RvGeo..40.1003G}, \doiurl{10.1029/2000RG000092}.
\bibitem[\protect\citeauthoryear{{Berger}}{2003}]{Berger03b}Groth, A., Ghil, M.: 2011, \pre{} \textbf{84}, 036206. \adsurl{2011PhRvE..84c6206G}, \doiurl{10.1103/PhysRevE.84.036206}.
\bibitem[\protect\citeauthoryear{{Berger}}{2003}]{Berger03b}Holmes, P., Lumley, J. L., Berkooz, G.: 1998, \textit{Turbulence, Coherent Structures, Dynamical Systems and Symmetry} Cambridge University Press, Cambridge, UK.\adsurl{1998tcsd.book.....H}
\bibitem[\protect\citeauthoryear{{Berger}}{2003}]{Berger03b}Howe, R., Christensen-Dalsgaard, J., Hill, F., Komm, R. W., Larsen, R. M., Schou, J., {\it et al.}: 2000, \textit{Science} \textbf{287}, 2456. \adsurl{2000Sci...287.2456H}, \doiurl{10.1126/science.287.5462.2456}.
\bibitem[\protect\citeauthoryear{{Berger}}{2003}]{Berger03b}Kane, R. P.: 2005, \solphys{} \textbf{227}, 155. \adsurl{2005SoPh..227..155K}, \doiurl{10.1007/s11207-005-1110-x}.
\bibitem[\protect\citeauthoryear{{Berger}}{2003}]{Berger03b}K{\"a}pyl{\"a}, M. J., K{\"a}pyl{\"a}, P. J., Olspert, N., Brandenburg, A., Warnecke, J., Karak, B. B., Pelt, J.	: 2016, \aap{} \textbf{589}, A56. \adsurl{2016A&A...589A..56K}, \doiurl{10.1051/0004-6361/201527002}.
\bibitem[\protect\citeauthoryear{{Berger}}{2003}]{Berger03b}Kato, C., Munakata, K., Yasue, S., Inoue, K., McDonald, F. B.: 2003, \jgr{} \textbf{108}, 1367. \adsurl{2003JGRA..108.1367K}, \doiurl{10.1029/2003JA009897}.
\bibitem[\protect\citeauthoryear{{Berger}}{2003}]{Berger03b}Knaack, R., Stenflo, J. O., Berdyugina, S. V.: 2004, \aap{} \textbf{418}, 17. \adsurl{2004A&A...418L..17K}, \doiurl{10.1051/0004-6361:20040107}.
\bibitem[\protect\citeauthoryear{{Berger}}{2003}]{Berger03b}Krivova, N. A., Solanki, S. K.: 2002, \aap{} \textbf{394}, 701. \adsurl{2002A&A...394..701K}, \doiurl{10.1051/0004-6361:20021063}.
\bibitem[\protect\citeauthoryear{{Berger}}{2003}]{Berger03b}Laurenza, M., Vecchio, A., Storini, M., Carbone, V.: 2012, \apj{} \textbf{749}, 167. \adsurl{2012ApJ...754..155L}, \doiurl{10.1088/0004-637X/754/2/155}.
\bibitem[\protect\citeauthoryear{{Berger}}{2003}]{Berger03b}Lockwood, M.: 2001, \textit{J. Geophys. Res.} \textbf{106}, 16021. \adsurl{2001JGR...10616021L}, \doiurl{10.1029/2000JA000115}.
\bibitem[\protect\citeauthoryear{{Berger}}{2003}]{Berger03b}Lou, Y. Q.: 2000, \apj{} \textbf{540}, 1102. \adsurl{2000ApJ...540.1102L}, \doiurl{10.1086/309387}.
\bibitem[\protect\citeauthoryear{{Berger}}{2003}]{Berger03b}Mancuso, S., Raymond, J. C.: 2015, \aap{} \textbf{573}, A33.  \adsurl{2015A&A...573A..33M}, \doiurl{10.1051/0004-6361/201424898}.
\bibitem[\protect\citeauthoryear{{Berger}}{2003}]{Berger03b}Mancuso, S., Raymond, J. C., Rubinetti, S., Taricco, C.: 2016, \aap{} \textbf{592}, L8. \adsurl{2016A&A...592L...8M}, \doiurl{10.1051/0004-6361/201628769}.
\bibitem[\protect\citeauthoryear{{Berger}}{2003}]{Berger03b}Mann, M. E., Lees, J. M.: 1996. \textit{Clim. Change} \textbf{33}, 409. \adsurl{1996ClCh...33..409M}, \doiurl{10.1007/BF00142586}.
\bibitem[\protect\citeauthoryear{{Berger}}{2003}]{Berger03b}Metcalfe, T. S., Buccino, A. P., Brown, B. P., Mathur, S., Soderblom, D. R., Henry, T. J., {\it et al.}: 2013, \apjl{} \textbf{763}, L26. \adsurl{2013ApJ...763L..26M}, \doiurl{10.1088/2041-8205/763/2/L26}.
\bibitem[\protect\citeauthoryear{{Berger}}{2003}]{Berger03b}Mursula, K., Usoskin, I., Zieger, B.: 1997, \solphys{} \textbf{176}, 201. \adsurl{1997SoPh..176..201M}, \doiurl{10.1023/A:1004982203293}.
\bibitem[\protect\citeauthoryear{{Berger}}{2003}]{Berger03b}Obridko, V. N., Badalyan, O. G.: 2014, \textit{Astron. Rep.} \textbf{58}, 936. \adsurl{2014ARep...58..936O}, \doiurl{10.1134/S1063772914120075}.
\bibitem[\protect\citeauthoryear{{Berger}}{2003}]{Berger03b}Obridko, V. N., Shelting, B. D.: 2001, \textit{Astron. Rep.} \textbf{45}, 1012. \adsurl{2001ARep...45.1012O}, \doiurl{10.1134/1.1426132}.
\bibitem[\protect\citeauthoryear{{Berger}}{2003}]{Berger03b}Ol{\'a}h, K., Koll{\'a}th, Z., Granzer, T., Strassmeier, K. G., Lanza, A. F., J{\"a}rvinen, S.,{\it et al.}: 2009, \aap{} \textbf{501}, 703. \adsurl{2009A&A...501..703O}, \doiurl{10.1051/0004-6361/200811304}.
\bibitem[\protect\citeauthoryear{{Berger}}{2003}]{Berger03b}Paularena, K. I., Szabo, A., Richardson, J. D.: 1995, \grl{} \textbf{22}, 3001. \adsurl{1995GeoRL..22.3001P}, \doiurl{10.1029/95GL02802}.
\bibitem[\protect\citeauthoryear{{Berger}}{2003}]{Berger03b}Plaut, G., Vautard, R.: 1994, \textit{J. Atmos. Sci.} \textbf{51} 210. \adsurl{1994JAtS...51..210P}, \doiurl{10.1175/1520-0469(1994)051<0210:SOLFOA>2.0.CO;2}.
\bibitem[\protect\citeauthoryear{{Berger}}{2003}]{Berger03b}Polygiannakis, J., Preka-Papadema, P., Moussas, X.: 2003, \mnras{} \textbf{343}, 725. \adsurl{2003MNRAS.343..725P}, \doiurl{10.1046/j.1365-8711.2003.06705.x}.
\bibitem[\protect\citeauthoryear{{Berger}}{2003}]{Berger03b}Richardson, J. D., Paularena, K. I., Belcher, J. W., Lazarus, A. J.: 1994, \grl{} \textbf{21}, 1559. \adsurl{1994GeoRL..21.1559R}, \doiurl{10.1029/94GL01076}.
\bibitem[\protect\citeauthoryear{{Berger}}{2003}]{Berger03b}Rieger, E., Share, G. H., Forrest, D. J., Kanbach, G., Reppin, C., Chupp, E. L.: 1984, \nat{} \textbf{312}, 623. \adsurl{1984Natur.312..623R}, \doiurl{10.1038/312623a0}.
\bibitem[\protect\citeauthoryear{{Berger}}{2003}]{Berger03b}Ryb{\'a}k, J., Antalov{\'a}, A., Storini, M.: 2001, \ssr{} \textbf{97}, 359. \adsurl{2001SSRv...97..359R}, \doiurl{10.1023/A:1011805923567}.
\bibitem[\protect\citeauthoryear{{Berger}}{2003}]{Berger03b}Rybansky, M., Rusin, V., Minarovjech, M., Gaspar, P.: 1994, \solphys{} \textbf{152}, 153. \adsurl{1994SoPh..152..153R}, \doiurl{10.1007/BF01473198}.
\bibitem[\protect\citeauthoryear{{Berger}}{2003}]{Berger03b}Shirai, T.: 2004, \solphys{} \textbf{222}, 199. \adsurl{2004SoPh..222..199S}, \doiurl{10.1023/B:SOLA.0000043565.83411.ec}.
\bibitem[\protect\citeauthoryear{{Berger}}{2003}]{Berger03b}Simard, C., Charbonneau, P., Bouchat, A.: 2013,\apj{} \textbf{768}, 16. \adsurl{2013ApJ...768...16S}, \doiurl{10.1088/0004-637X/768/1/16}.
\bibitem[\protect\citeauthoryear{{Berger}}{2003}]{Berger03b}Sturrock, P. A., Bush, R., Gough, D. O., Scargle, J. D.: 2015, \apj{} \textbf{804}, 47. \adsurl{2015ApJ...804...47S}, \doiurl{10.1088/0004-637X/804/1/47}.
\bibitem[\protect\citeauthoryear{{Berger}}{2003}]{Berger03b}Szabo, A., Lepping, R. P., King, J. H.: 1995, \grl{} \textbf{22}, 1845. \adsurl{1995GeoRL..22.1845S}, \doiurl{10.1029/95GL01737}.
\bibitem[\protect\citeauthoryear{{Berger}}{2003}]{Berger03b}Taricco, C., Mancuso, S., Ljungqvist, F. C., Alessio, S., Ghil, M.: 2015a, \textit{Clim. Dyn.} \textbf{45}, 83. \adsurl{2015ClDy...45...83T}, \doiurl{10.1007/s00382-014-2331-1}.
\bibitem[\protect\citeauthoryear{{Berger}}{2003}]{Berger03b}Taricco, C., Vivaldo, G., Alessio, S., Rubinetti, S., Mancuso, S.: 2015b, \textit{Clim. of the Past} \textbf{11}, 509.  \adsurl{2015CliPa..11..509T}, \doiurl{10.5194/cp-11-509-2015}.
\bibitem[\protect\citeauthoryear{{Berger}}{2003}]{Berger03b}Taricco, C., Alessio, S., Rubinetti, S., Vivaldo, G., Mancuso, S.: 2016, \textit{Nature Sc. Data} \textbf{3}, 160042. \adsurl{2016NatSD...360042T}, \doiurl{10.1038/sdata.2016.42}.
\bibitem[\protect\citeauthoryear{{Berger}}{2003}]{Berger03b}Thomson, D. J.: 1982, \textit{Proc IEEE} \textbf{70}, 1055. \adsurl{1982IEEEP..70.1055T}.
\bibitem[\protect\citeauthoryear{{Berger}}{2003}]{Berger03b}Ulrich, R. K., Tran, T.: 2013, \apjl{} \textbf{768}, 189. \adsurl{2013ApJ...768..189U}, \doiurl{10.1088/0004-637X/768/2/189}.
\bibitem[\protect\citeauthoryear{{Berger}}{2003}]{Berger03b}Vautard, R., Ghil, M.: 1989, \textit{Physica D} \textbf{35}, 395. \adsurl{1989PhyD...35..395V}, \doiurl{10.1016/0167-2789(89)90077-8}.
\bibitem[\protect\citeauthoryear{{Berger}}{2003}]{Berger03b}Vecchio, A., Carbone, V.: 2008, \apjl{} \textbf{683}, 536. \adsurl{2008ApJ...683..536V}, \doiurl{10.1086/589768}.
\bibitem[\protect\citeauthoryear{{Berger}}{2003}]{Berger03b}Vecchio, A., Carbone, V.: 2009, \aap{} \textbf{502}, 981. \adsurl{2009A&A...502..981V}, \doiurl{10.1051/0004-6361/200811024}.
\bibitem[\protect\citeauthoryear{{Berger}}{2003}]{Berger03b}Vecchio, A., Laurenza, M., Carbone, V., Storini, M.: 2010, \apjl{} \textbf{709}, 1. \adsurl{2010ApJ...709L...1V}, \doiurl{10.1088/2041-8205/709/1/L1}.
\bibitem[\protect\citeauthoryear{{Berger}}{2003}]{Berger03b}Vecchio, A., Laurenza, M., Meduri, D., Carbone, V., Storini, M.: 2012, \apjl{} \textbf{749}, 27. \adsurl{2012ApJ...749...27V}, \doiurl{10.1088/0004-637X/749/1/27}.
\bibitem[\protect\citeauthoryear{{Berger}}{2003}]{Berger03b}Wang, Y.-M., \& Sheeley, N. R., Jr.: 2003, \apj{} \textbf{590}, 1111. \adsurl{2003ApJ...590.1111W}, \doiurl{10.1086/375026}.
\bibitem[\protect\citeauthoryear{{Berger}}{2003}]{Berger03b}Wolff, C. L.: 1992, \solphys{} \textbf{142}, 187. \adsurl{1992SoPh..142..187W}, \doiurl{10.1007/BF00156641}.
\bibitem[\protect\citeauthoryear{{Berger}}{2003}]{Berger03b}Zaqarashvili, T. V., Carbonell, M., Oliver, R., Ballester, J. L.: 2010, \apj{} \textbf{709}, 749. \adsurl{2010ApJ...709..749Z}, \doiurl{10.1088/0004-637X/709/2/749}.

\end{thebibliography}

\end{article} 
\end{document}